\documentclass[11pt]{IEEEtran}
\IEEEoverridecommandlockouts
\usepackage[utf8]{inputenc}
\usepackage[tmargin=1in, lmargin=1.25in, rmargin=1.25in, bmargin=1in]{geometry}
\usepackage[utf8]{inputenc}
\usepackage[utf8]{inputenc} 
\usepackage[T1]{fontenc}
\usepackage{url}
\usepackage{ifthen}
\usepackage{enumitem}
\usepackage{tabularx}
\usepackage{cite}
\usepackage{bbm, dsfont}
\usepackage{dblfloatfix}
\usepackage[cmex10]{amsmath} 

\usepackage[percent]{overpic}
\usepackage{cite}
\usepackage[most]{tcolorbox}
\usepackage{amsmath,amssymb,amsfonts}
\usepackage{mathtools}
\usepackage{amsthm}
\usepackage{algorithmic}
\usepackage{graphicx}
\usepackage{textcomp}
\usepackage{tikz}
\usepackage{hyperref}
\usepackage{caption}
\usepackage{cuted}
\usepackage{romannum}
\usepackage[utf8]{inputenc}
\usepackage{float}
\usepackage{pgfplots} 
\usepackage{pgfgantt}
\usepackage{pdflscape}
\usepackage{amssymb}
\usepackage{comment}
\usepackage{pst-plot}
\usetikzlibrary{spy}
\usetikzlibrary{positioning,calc}
\usetikzlibrary{decorations.pathmorphing,calc,shapes,shapes.geometric,patterns}
\usetikzlibrary{shapes.multipart}
\usepackage{tikz}
\usetikzlibrary{arrows.meta, positioning, shapes.geometric}
\usepackage{xfrac}
\usepackage{colortbl}
\usepackage{cancel}
\usepackage{smartdiagram}

\usetikzlibrary{arrows,positioning,calc,intersections}
\usetikzlibrary{datavisualization.formats.functions}
\def\BibTeX{{\rm B\kern-.05em{\sc i\kern-.025em b}\kern-.08em
    T\kern-.1667em\lower.7ex\hbox{E}\kern-.125emX}}
    
\usepackage{pgfplots}
\usepgfplotslibrary{fillbetween}
\usetikzlibrary{arrows, decorations.markings}
\usepackage{fontawesome5}

\newtheorem{example}{Example}

\definecolor{mutedteal}{RGB}{31,148,148}

\definecolor{calpolypomonagreen}{rgb}{0.12, 0.3, 0.17}
\definecolor{green4}{RGB}{0,170,0}
\newcounter{remarkcount}

\newcommand{\circlearrow}{}
\DeclareRobustCommand{\circlearrow}{%
  \mathrel{\vphantom{\rightarrow}\mathpalette\circle@arrow\relax}%
}
\newcommand{\circle@arrow}[2]{%
  \m@th
  \ooalign{%
    \hidewidth$#1\circ\mkern1mu$\hidewidth\cr
    $#1-$\cr}%
}
\makeatother

\usepackage{amsmath, amssymb, amsfonts, amsthm}
\usepackage{bm}
\usepackage{xcolor}
\usepackage{framed}
\usepackage{lipsum}
\definecolor{brightlavender}{rgb}{0.75, 0.58, 0.89}
\usepackage{pgf}
\usepackage{tikz}
\usepackage{multicol}
\usetikzlibrary{arrows,automata}
\usetikzlibrary{positioning}
\usetikzlibrary{decorations.pathmorphing}
\usetikzlibrary{shapes.geometric}
\usetikzlibrary{fit}
\usetikzlibrary{backgrounds}
\definecolor{UTFPRYellow}{RGB}{254, 203, 41}
\definecolor{hollywoodcerise}{rgb}{0.96, 0.0, 0.63}
\tikzset{planet/.append style={
 draw=UTFPRYellow,thick, fill=UTFPRYellow!40
 },
 satellite/.append style={
 draw=hollywoodcerise, fill=brightlavender!50
 },
 }
\usepackage{hhline}
\usepackage[caption=false,font=footnotesize]{subfig}
\usepackage{svg}
\usepackage{hyperref}

\definecolor{mutedteal}{rgb}{0.31, 0.55, 0.47}  
\theoremstyle{definition}

\theoremstyle{remark}

\DeclarePairedDelimiterX{\infdivx}[2]{(}{)}{%
  #1\;\delimsize\|\;#2%
}

\begin{document}
\onecolumn
\title{Common Randomness: A Key Enabler of Trustworthy 6G
Communication Systems
}
\author{
\IEEEauthorblockN{Rami Ezzine\IEEEauthorrefmark{1}\IEEEauthorrefmark{4}, Moritz Wiese\IEEEauthorrefmark{1}\IEEEauthorrefmark{4}, Wafa Labidi\IEEEauthorrefmark{1}\IEEEauthorrefmark{4}, Christian Deppe\IEEEauthorrefmark{2}\IEEEauthorrefmark{4} and Holger Boche\IEEEauthorrefmark{1}\IEEEauthorrefmark{3}\IEEEauthorrefmark{4}\IEEEauthorrefmark{5}\IEEEauthorrefmark{6}}

\IEEEauthorblockA{\IEEEauthorrefmark{1}Technical University of Munich, Munich, Germany\\
\IEEEauthorrefmark{2}Technische Universit\"at Braunschweig, Brunswick, Germany\\
\IEEEauthorrefmark{3}\color{black}Cluster of Excellence ”Centre for
Tactile Internet with Human-in-the-Loop” (CeTI) of the Technische Universit\"at
Dresden, Dresden, Germany\color{black}\\
\IEEEauthorrefmark{4}BMFTR Research Hub 6G-life, Germany\\
\IEEEauthorrefmark{5}Munich Center for Quantum Science and Technology (MCQST) \\
\IEEEauthorrefmark{6}Munich Quantum Valley (MQV) \\
Email: rami.ezzine@tum.de, wiese@tum.de, wafa.labidi@tum.de, christian.deppe@tu-bs.de and boche@tum.de }}

\maketitle
\thispagestyle{plain}
\pagenumbering{arabic}
\pagestyle{plain}
\begin{abstract}
Common randomness (CR) is a valuable resource for enhancing the trustworthiness of 6G communication systems. This article highlights the role of CR in improving scalability, security, and resilience in distributed 6G architectures. It provides an overview of information-theoretic two-source models for CR generation and summarizes known capacity results, together with their implications for practical systems under different communication settings, including one-way, two-way, helper-assisted, and interactive scenarios. Furthermore, the article explores CR-assisted secure identification over Gaussian channels as an application and highlights selected open research problems in CR generation for next-generation wireless systems. 

The article targets researchers and practitioners interested in bridging information-theoretic tools with emerging 6G communication systems.
\end{abstract}
\section{Introduction}

\begin{figure*}[t]
    \centering
    \input{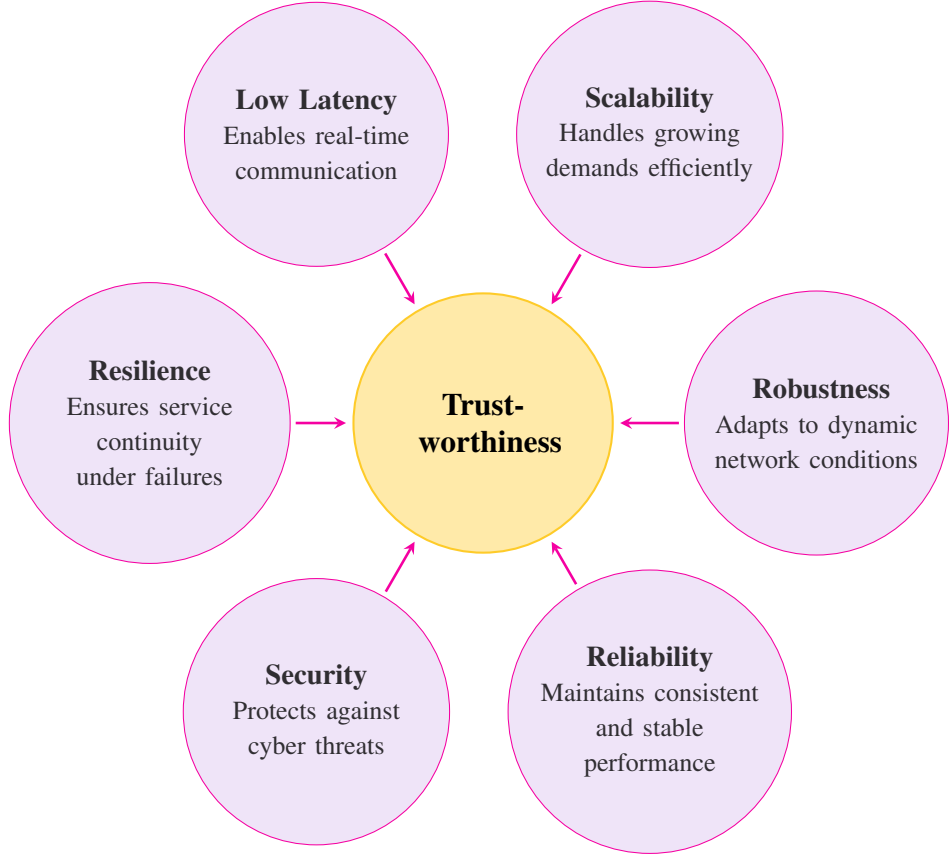}
    \caption{Key aspects contributing to trustworthiness in 6G communication systems.}
    \label{figtrustworthiness}
\end{figure*}

\color{black}
The evolution of 6G mobile networks brings new challenges, particularly in ensuring trustworthiness, i.e., the ability of a system to operate reliably, securely, and consistently under varying conditions \cite{6Gandtrustworthiness}. Modern 6G communication systems must satisfy several requirements to ensure trustworthiness, including low latency, high reliability, scalability, robust security, and strong resilience, as illustrated in Fig.~\ref{figtrustworthiness}. 
Achieving these goals requires communication paradigms that extend beyond traditional frameworks. A promising tool to address these challenges is \textbf{common randomness (CR)}. 
\begin{tcolorbox}
[
    title=What is CR?,
    colback=blue!5,
    colframe=blue!40!black,
    coltitle=white,
    colbacktitle=blue!40!black,
    boxrule=0.5pt,
    arc=1mm
]
\color{black}CR refers to a shared random variable (RV) that multiple parties agree on with high probability despite limited communication.
\end{tcolorbox}
\begin{center}
\fbox{%
\begin{minipage}{0.95\columnwidth}
\small
\textit{Accepted for publication in IEEE BITS the Information Theory Magazine.}

\vspace{2pt}

\textcopyright\ 2026 IEEE. Personal use of this material is permitted.
Permission from IEEE must be obtained for all other uses, in any current or
future media, including reprinting/republishing this material for advertising
or promotional purposes, creating new collective works, for resale or
redistribution to servers or lists, or reuse of any copyrighted component of
this work in other works. 

DOI: 10.1109/MBITS.2026.3735678
\end{minipage}%
}
\end{center}
As a fundamental resource in Post-Shannon theory, extending classical information theory toward emerging communication needs \cite{6GpostShannon}, CR supports secure coordination, synchronization, identification, and resilience in distributed 6G architectures. \color{black}
CR also arises in quantum communication, where it can be distilled from quantum states \cite{q5}. In this article, quantum CR is mentioned only to provide broader context; the analysis and open problems focus on classical CR generation in distributed 6G systems.

\color{black}Decentralized methods provide an efficient approach to generating CR, where network nodes exploit shared physical properties, such as wireless channel reciprocity, to locally extract the same random values. This approach reduces communication overhead, avoids single points of failure, and offers greater scalability and flexibility, making it well-suited to distributed 6G communication systems.

To provide a systems perspective, this article uses cooperative vehicle-to-vehicle (V2V) communication as a running example to illustrate how CR can support trustworthiness in distributed 6G communication systems. This scenario is revisited across different discussions. 

\color{black}Building on prior works addressing individual CR techniques and scenarios, this article synthesizes established information-theoretic results on CR and connects them to 6G applications and trustworthiness through a unified systems perspective. Its contribution lies in providing a connective narrative and systems-level interpretation of these results within a common trustworthiness framework. \color{black} After reading this article, the reader will gain insights into:
\begin{itemize}
    \item why CR is fundamental to trustworthiness in distributed 6G networks;
    \item the main information-theoretic models for CR generation under different communication \color{black}settings; \color{black}
    \item how CR enables practical applications, such as secure identification; and
    \item \color{black}selected \color{black} open problems in CR generation for next-generation wireless systems.
\end{itemize}

\textbf{Outline:}
The remainder of this article is organized as follows. Section~II discusses the role of CR in enhancing the trustworthiness of 6G communication systems from a system-level perspective. Section~III reviews the main information-theoretic models for CR generation and summarizes their implications for practical systems. Section~IV presents CR-assisted secure identification over Gaussian channels as a representative application, and Section~V discusses open research challenges and future directions.

\textbf{Notation:} 
Throughout the article, RVs are denoted by uppercase letters, their realizations by lowercase letters, and sequences of length $n$ by superscripts (e.g., $X^n$). For any discrete RV $X$, $H(X)$ denotes its entropy. For any RVs $X$ and $Y$, $I(X;Y)$ denotes the mutual information between $X$ and $Y.$ The binary entropy function is denoted by $H_2(\cdot)$. The notation $X \circlearrow{Y} \circlearrow{Z}$ denotes a Markov chain.
\color{black}

\begin{figure*}
\centering
\input{ID_scheme}
\caption{\color{black}Application of identification in V2V communication (\small Traffic image adapted from vecstock, Freepik,
free use with attribution,
\href{https://www.magnific.com/ai/docs/licenses-attribution}{license and attribution},
\href{https://www.freepik.com/free-ai-image/speeding-cars-blur-blue-modern-city-rush-generated-by-ai_41148241.htm}{source}).}
\label{id}
\end{figure*}
\section{The Role of CR in Future 6G Communication Systems}
\label{roleofCR}
\color{black}

This section is organized around three trustworthiness aspects highlighted in Fig.~\ref{figtrustworthiness}: scalability, resilience, and security, and discusses how CR contributes to each of them.

\subsection{Meeting the Scalability Requirements with CR}

\color{black}As highlighted in Fig.~\ref{figtrustworthiness}, future 6G systems require communication methods with improved scaling behavior in terms of energy consumption and hardware requirements. Improved scaling is needed to avoid situations in which the data generation rate grows faster than the transmission rate \cite{angetal}. \color{black}
\color{black} \color{black} This
calls for communication paradigms that reduce processing and transmission
overhead while still meeting performance and reliability demands.

Identification \cite{identification} is an example of such methods. It is applied in alarm and control systems \cite{id1,id2}, and more recently also in machine learning for semantic communication \cite{id3}. 
\begin{tcolorbox}
[
    title=What is Identification?,
    colback=blue!5,
    colframe=blue!40!black,
    coltitle=white,
    colbacktitle=blue!40!black,
    boxrule=0.5pt,
    arc=1mm,
]
Identification is a communication paradigm where the receiver only determines whether a specific message of interest was sent, rather than recovering the entire transmitted message as in the classical Shannon transmission scheme. 
\end{tcolorbox}
\color{black}An example of the scalability challenge is found in vehicle-to-vehicle (V2V) communication: \color{black} many sensors do not transmit all generated data since only part of the information is relevant to other vehicles. For instance, Vehicle B may need to know whether a specific message about the future movement of Vehicle A has been sent. A classical Shannon-based scheme would require Vehicle B to receive and fully decode all messages from Vehicle A, including those that are not of interest to Vehicle B. This leads to unnecessary energy consumption, hardware complexity, and limits scalability as the number of messages and participants increases. Identification coding offers a more efficient alternative by allowing Vehicle B to determine, without full decoding, whether the relevant message has been sent, as illustrated in Fig.~\ref{id}.

CR plays a crucial role by enhancing the identification capacity of communication channels~\cite{Generaltheory}. In fact, many known random identification code constructions begin with generating CR as a first step. Operationally, the generated CR is used to construct shared random codebooks available to both the transmitter and receiver. If CR is already available, only a few bits need to be transmitted to identify the message of interest, significantly reducing communication overhead while increasing the achievable identification performance. 

A key reason for this scalability advantage is that identification coding enables a double-exponential growth in the number of identifiable messages with blocklength, far exceeding the exponential growth achievable in classical transmission, as shown by Ahlswede and Dueck~\cite{identification}. This scaling behavior highlights the potential of identification schemes to support future 6G applications by reducing communication overhead in large-scale systems.

Another application of CR generation is in source coding, where it improves performance in output-constrained lossy scenarios. As shown in \cite{applicationsourcecoding}, shared randomness enables more efficient rate-distortion trade-offs and simplifies encoding and decoding. In 6G applications, including cooperative V2V systems, real-time video, and tactile Internet, these benefits can support scalable communication with reduced resource consumption.

\subsection{Meeting the Resilience Requirements with CR}

The resilience aspect highlighted in Fig.~\ref{figtrustworthiness} motivates the development of communication methods that can maintain reliable operation under adversarial conditions and changing environments. CR can be exploited to achieve resilience by design for tactile Internet and quantum communication systems, which are key components of 6G networks. When legitimate parties have access to a common random source for coordination, communication becomes resilient to denial-of-service (\color{black}DOS\color{black}) attacks from a jammer, \color{black}requiring only a few bits of CR to mitigate jamming attacks \cite{jammerref}. \color{black}

\color{black}A well-known information-theoretic model for communication in the presence of a jammer is the arbitrarily varying channel (AVC). \color{black}

\begin{tcolorbox}
[
    title=What is an AVC?,
    colback=blue!5,
    colframe=blue!40!black,
    coltitle=white,
    colbacktitle=blue!40!black,
    boxrule=0.5pt,
    arc=1mm
]
An AVC models a communication system with arbitrarily varying channel states, capturing scenarios such as interference, jamming, and rapidly changing wireless conditions while requiring robustness to worst-case variations.
\end{tcolorbox}

\color{black}CR plays a significant role in improving coding over the AVC \cite{jammerref}. \color{black} In this model, a jammer attempts to degrade communication by selecting the worst possible state to prevent reliable message transmission, as depicted \color{black}in Fig. \ref{jammer}. \color{black}However, the availability of CR allows the legitimate parties to randomize their encoding and decoding strategies, effectively transforming these attacks into noise and preventing a communication breakdown caused by the jamming attack. This enables the legitimate parties to continue communication and even initiate countermeasures against the jammer \cite{jammerref}.

\begin{figure*} 
\centering 
\input{Model_CR_jammer} 
\caption{System model of communication under the adversarial attack of a jammer (\small Alice and Bob icons adapted from Tomasz ``odder'' Kozlowski,
CC BY-SA 3.0,
\href{https://creativecommons.org/licenses/by-sa/3.0/}{license},
via Wikimedia Commons,
\href{https://commons.wikimedia.org/wiki/File:Asymmetric_cryptography_-_step_1.svg}{source};
Jammer icon taken from the GNOME Project,
CC BY-SA 3.0 US,
\href{https://creativecommons.org/licenses/by-sa/3.0/us/}{license},
via Wikimedia Commons,
\href{https://commons.wikimedia.org/wiki/File:Gnome3-devilish.svg}{source}).} 
\label{jammer} 
\end{figure*}

\color{black}While it has been recognized that CR can enhance message transmission capacity in certain scenarios, its potential has been largely overlooked in current communication systems. In particular, CR is crucial for maintaining reliable communication in the presence of powerful jamming attacks, where conventional methods may fail. More generally, CR improves system robustness and performance by enabling stronger protection against interference and adversarial behavior. CR-based randomization at the physical layer complements existing techniques, such as frequency hopping and cryptographic methods, by providing additional resilience against adversarial actions. 

\color{black}Another interesting application of CR generation is in Integrated Sensing and Communication (ISAC), an important and innovative key feature of 6G systems \cite{6GPerspective}. By combining sensing and communication, ISAC allows adaptive and efficient applications, but it also exposes systems to changing channel conditions and potential disruptions \cite{6Gandtrustworthiness}. CR improves ISAC's resilience by helping systems operate reliably, even under fluctuating or imperfect conditions.

Recent work \cite{distributionpreserving} has shown that CR enables error-free state estimation and accurate reconstruction of state distributions in joint sensing and communication applications. \color{black}CR helps align the way different nodes interpret noisy measurements, making joint sensing and communication more robust. \color{black} This allows ISAC systems to maintain reliable coordination even with incomplete or noisy data, ensuring robust operation in unpredictable environments. 
\subsection{Meeting the Security Requirements with CR}

The security aspect highlighted in Fig.~\ref{figtrustworthiness} motivates the development of communication methods that provide protection against eavesdropping and other security threats. Securing future communication systems based on physical-layer security is a promising method to withstand complex attacks, including those carried out with quantum computers, and represents an important research direction. Physical-layer security provides a foundation for long-term protection that does not rely on computational assumptions. Moreover, it is a valuable feature for 6G communication systems, offering an added layer of defense against eavesdropping and security threats.

\begin{figure*}[t]
\centering
\input{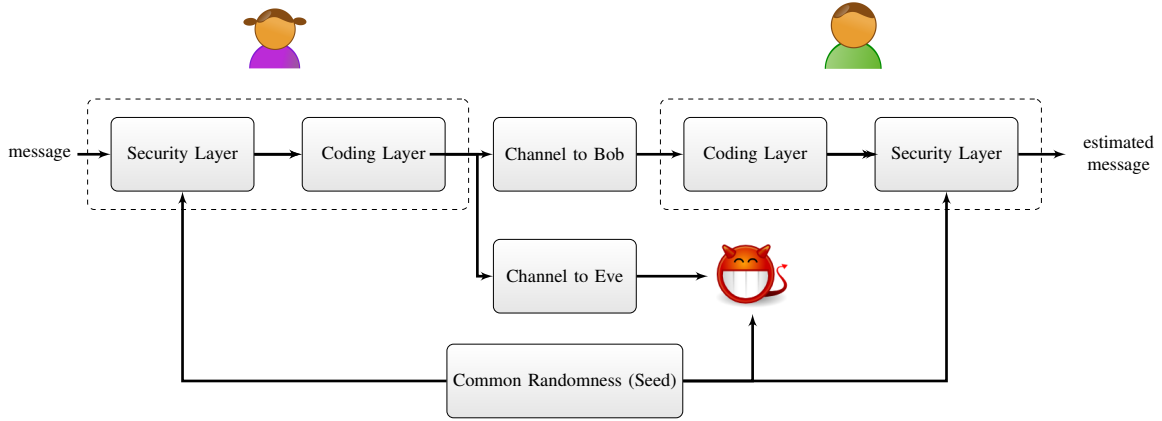}
\caption{Seeded modular coding scheme (\small Alice and Bob icons: see attribution in Fig. \ref{jammer}; \small
Eve icon taken from the Tango Desktop Project,
released into the public domain,
via Wikimedia Commons,
\href{https://commons.wikimedia.org/wiki/File:Face-devil-grin.svg}{source}).}
\label{seededmodularcoding}
\end{figure*}

CR plays a significant role in enabling security in modular coding schemes for communication. Modular schemes for semantic security have been developed in \cite{semanticwiretap}\color{black}, and they work efficiently for all practically relevant channels. A typical scenario in modular coding is when legitimate parties have access to CR as a resource. This CR can be used as a seed \cite{semanticsecurity}, as depicted in Fig. \ref{seededmodularcoding}. 

\begin{tcolorbox}
[
    title=What is a seed?,
    colback=blue!5,
    colframe=blue!40!black,
    coltitle=white,
    colbacktitle=blue!40!black,
    boxrule=0.5pt,
    arc=1mm
]
\color{black}A seed is a shared piece of randomness that serves as a basis for generating cryptographic keys, enhancing encryption protocols, or enabling secure communication.\color{black}
\end{tcolorbox} In particular, the CR seed allows standard reliability-oriented channel codes to be transformed into semantic-security codes by introducing additional randomization that prevents an adversary from exploiting the transmitted code structure. It is essential for ensuring reliable transmission, protecting against attacks, and maintaining robustness in security schemes, particularly in the presence of evolving threats such as quantum computers.

\color{black}\color{black}Beyond modular coding, \color{black} CR is also relevant in the secret key generation problem from correlated observations. \color{black}Under additional secrecy constraints, the generated CR can be used for privacy amplification to extract a secret key, as demonstrated in foundational works \cite{part1,Maurer}. The generated secret keys enable various cryptographic tasks, including secure transmission and authentication. 

\color{black}CR can also enhance secure message identification. For Gaussian channels with positive secrecy capacity, CR-assisted schemes may achieve higher secure identification rates than those relying only on local randomness, with secrecy in identification
obtained for free. A detailed discussion is provided in Section~\ref{application}. \color{black}

\color{black}
\section{Information-Theoretic Models for Common Randomness Generation and Their Fundamental Limits}
\label{inftheoreticmodel}
\color{black}Information-theoretic models for decentralized CR generation provide fundamental abstractions for distributed 6G systems. Operationally, the CR capacity of such models characterizes the maximum achievable CR rate under given source and communication constraints. Several two-source models have been studied in the literature, as discussed next.
\subsection{Two-Source Model with One-Way Communication} 
\label{twosourcemodelsectiononeway}
\begin{figure}[H]
    \centering
    \input{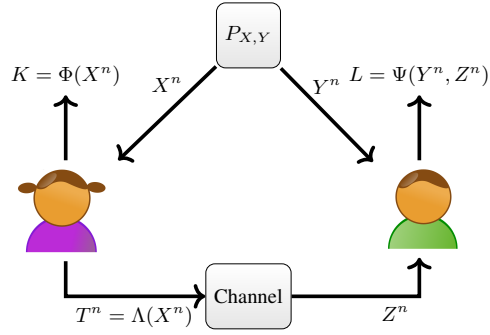}
\caption{Two-source model for CR generation with one-way communication (\small Alice and Bob icons: see attribution in Fig. \ref{jammer}).}
    \label{fig:codingscheme Protocol}
\end{figure}

Consider a memoryless multiple-source (MMS), which has two components, represented by the generic variables \(X\) and \(Y\), and governed by the probability law \(P_{X,Y}\). The MMS generates independent and identically distributed (i.i.d.) samples of \((X,Y)\). The outputs of \(X\) are observed only by Alice, and those of \(Y\) only by Bob. The joint distribution of \((X,Y)\) is known to both parties. In addition, Alice can send information to Bob over some channel. No other resources are available to Alice or Bob, as depicted in Fig.~\ref{fig:codingscheme Protocol}.


A CR-generation protocol of block-length $n$ consists of:
\begin{enumerate}[label=(\alph*)]
    \item A function $\Phi$ that maps $X^n$ into a RV $K$ generated by Alice.
    \item A function $\Lambda$ that maps $X^n$ into the channel input sequence generated by Alice.
    \item A function $\Psi$ that maps $Y^n$ and the channel output sequence $Z^n$ into a RV $L$ generated by Bob.
\end{enumerate}
Such a protocol induces a pair of RVs $(K,L)$ whose joint distribution is determined by $P_{X,Y}$ and by the channel.

The (uniform) CR capacity is the maximum rate \( \frac{H(K)}{n} \) at which Alice and Bob can generate (uniform or nearly uniform) shared RVs with vanishing disagreement probability as \(n\) grows. For discrete MMSs, it is upper-bounded by \(H(X)\). In the absence of communication, the CR capacity equals the maximum entropy of a common function of \(X\) and \(Y\) \cite{CIT-122}, and is zero for indecomposable joint distributions. For rate-limited noiseless channels or DMCs, CR  capacity formulas were derived in \cite{part2}.

\begin{tcolorbox}[
    title=CR Capacity Characterization,
    colback=blue!5,
    colframe=blue!40!black,
    coltitle=white,
    colbacktitle=blue!40!black,
    boxrule=0.5pt,
    arc=1mm
]

For a noiseless channel with rate constraint $R$, the CR capacity is characterized by
\[
L_f^{X,Y}(R)
=
\max_{\substack{
U\in\mathcal U_f\\
U\circlearrow{X}\circlearrow{Y}\\
I(U;X)-I(U;Y)\leq R
}}
I(U;X),
\]
where $\mathcal U_f$ is a well-defined set of finite-valued RVs.
\end{tcolorbox}
The RV $U$ serves as a single-letter representation of the CR generated by Alice, based solely on her observation of the generic variable \( X \).  
\color{black}Under the Markovity constraint \( U \circlearrow{X} \circlearrow{Y} \), \( I(U;X) - I(U;Y) \) represents the net information that Alice must send to Bob for him to reconstruct $U$. This difference shows how much of $U$ is already revealed by \( Y \): a larger value means less is known and more communication is needed, while a smaller or zero value means \( Y \) already provides most of the information about $U.$ 
The same formula applies to the second scenario, with the transmission capacity $R$ of the noiseless channel replaced by the transmission capacity of the DMC \cite{part2}.

\noindent \textbf{\color{black} Systems Perspective: \color{black}}
From a 6G perspective, this model represents a two-source network where two parties observe correlated wireless measurements or sensing data and use one-way communication to establish shared randomness. For example, in cooperative V2V, the observations may correspond to channel state information, received signal measurements, or sensing data from a shared propagation environment, with one party sending reconciliation information to the other.
\color{black}
\begin{table*}[t] 
  \centering
  \arrayrulecolor[HTML]{000000}  
  \colorlet{headerblack}{blue!40!black}
 \colorlet{rowlightblack}{blue!5}
  \renewcommand{\arraystretch}{1.2}
  \footnotesize

  \begin{tabularx}{\textwidth}{|
    >{\arraybackslash}p{2.4cm}|
    >{\arraybackslash}p{3.9cm}|
    >{\arraybackslash}X|
  }
  \noalign{\hrule height 0.8pt} 
   \rowcolor{headerblack}
   \textcolor{white}{\textbf{Source Type}} &
   \textcolor{white}{\textbf{Channel Type}} &
   \textcolor{white}{\textbf{Capacity Results}} \\
  \noalign{\hrule height 0.8pt} 
  \noalign{\hrule height 0.8pt} 

  \noalign{\hrule height 0.8pt}
  \rowcolor{rowlightblack}
   finite single-state &
   multiple-antenna Gaussian with average power constraint &
   single-letter formula for the CR capacity \cite{optimalsignalprocessing} \\

  \noalign{\hrule height 0.8pt}
   finite single-state &
   multiple-antenna slow fading channels with arbitrary channel state distribution &
   single-letter bounds on the outage CR capacity \cite{journalpaperfading} \\

  \noalign{\hrule height 0.8pt}
  \rowcolor{rowlightblack}
   finite single-state &
   arbitrary &
   general formula for the UCR capacity and general bounds on the $\epsilon$-UCR capacity \cite{journalpaperCRarbitrary} \\

  \noalign{\hrule height 0.8pt}
   finite compound &
   rate-limited noiseless channel &
   single-letter compound CR capacity formula when the parties communicate and single-letter bounds on the compound CR capacity in the absence of communication \cite{compoundpaper} \\

  \noalign{\hrule height 0.8pt}
  \rowcolor{rowlightblack}
   sources with Polish alphabet &
   noisy memoryless &
   single-letter bounds on the CR capacity \cite{isit2025} \\

  \noalign{\hrule height 0.8pt} 
  \end{tabularx}

  \caption{Overview of the main results on CR capacity for several practical two-source models with one-way communication.}
  \label{summary}
\end{table*}

\color{black}CR generation aided by unidirectional communication has been investigated in the literature for several practical communication channels. We distinguish the following models:

\begin{itemize}
    \item \textbf{Memoryless
multiple-antenna additive Gaussian channels with average power constraint\cite{optimalsignalprocessing}:} \color{black}These channels are highly relevant to \textcolor{black}{6G wireless networks, supporting high data rates, low-latency communication, massive connectivity, and AI-driven services}.
\color{black}
\item \textbf{Multiple-antenna slow fading channels with arbitrary channel state distribution \cite{journalpaperfading}}: \color{black} These channels are particularly relevant for 6G systems, as they capture long-term, unpredictable variations in signal strength caused by environmental factors such as terrain, mobility, and obstacles, which are critical for applications like V2X communications, drone networks, and mobile edge computing.
\item \textbf{Arbitrary single-user channels \cite{journalpaperCRarbitrary}}: \color{black}For these channels, no assumptions are made regarding stationarity, ergodicity, or stability, making them well-suited for dynamic 6G environments that support various services, such as IoT deployments, real-time AR/VR, and ISAC applications.
\end{itemize}
\color{black}Beyond finite single-state sources, the following source categories have been investigated: \color{black}
\begin{itemize}
    \item \textbf{Finite compound sources \cite{compoundpaper}:} \color{black}They illustrate a practical scenario concerning source state \color{black} uncertainty\color{black}, where legitimate users lack precise knowledge of the actual source realization but know that the source belongs to a fixed uncertainty set. \color{black}This is particularly valuable for 6G networks to manage uncertainty in adaptive network management and resource allocation.
    \item \textbf{Sources defined on infinite Polish alphabet \cite{isit2025}:} They provide a robust framework for modeling continuous-valued data. In 6G systems, infinite Polish alphabets efficiently represent high-resolution sensor streams, environmental monitoring data, and large-scale IoT information.

\end{itemize}
\color{black}Across all cases, CR capacity formulas or bounds involve expressions similar to $L_f^{X,Y}$, adapted to the specific channel or source. For models involving infinite alphabets or continuous sources, the set \( \mathcal{U}_f \) is modified to reflect the corresponding source structure. \color{black}Overall, despite differences in the underlying communication models, the CR capacity is consistently characterized by optimizing shared information subject to the communication constraints. An overview of the CR capacity results and their corresponding references is provided in Table~\ref{summary}.\color{black}
\color{black}
\subsection{Two-Source Model with Two-Way Noiseless Communication}

\begin{figure}[htp!]
    \centering
    \input{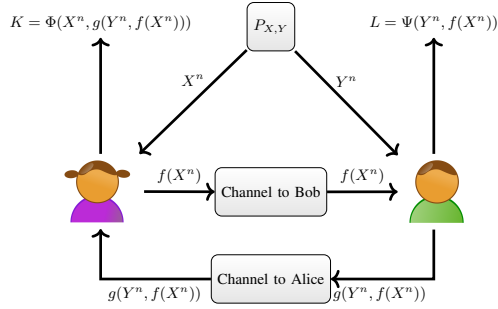}
    \caption{Two-source model for CR generation with two-way noiseless communication (\small Alice and Bob icons: see attribution in Fig. \ref{jammer}).}
    \label{CRtwowayfig}
\end{figure}

Consider the two-source CR generation problem with two-way noiseless communication, as 
depicted in Fig.~\ref{CRtwowayfig}.
For the discrete MMS described above, Alice observes $X^n$ and sends $f(X^n)$ to Bob over a noiseless channel of capacity $R_1>0$. Bob generates $L=\Psi(Y^n,f(X^n))$ and sends $g(Y^n,f(X^n))$ over a second noiseless channel of capacity $R_2>0$. Alice then generates $K=\Phi(X^n,g(Y^n))$. This model can be extended to multiple communication rounds \cite{survey}.

Compared with the one-way model, two-way communication enables Alice and Bob to iteratively refine the information extracted from their correlated observations. Both parties exchange messages, allowing improved agreement on CR.  The CR capacity expression is obtained by optimizing over auxiliary variables that represent the information extracted by both parties while satisfying the communication constraints in both directions \cite{part2}.

\noindent \textbf{\color{black} Systems Perspective:\color{black}} This model captures bidirectional message exchange between distributed nodes and is relevant to 6G scenarios supporting low-latency feedback, such as device-to-device communication, cooperative sensing, and distributed edge intelligence.
\subsection{Helper-Assisted Two-Source Model}

\begin{figure}[b!]
    \centering
    \input{model_with_helper}
    \caption{Helper-assisted two-source model for CR generation (\small Alice and Bob icons: see attribution in Fig. \ref{jammer}; \small Helper icon taken from Vincent Le Moign,
CC BY 4.0,
\href{https://creativecommons.org/licenses/by/4.0/}{license},
via Wikimedia Commons,
\href{https://commons.wikimedia.org/wiki/File:133-man-judge-1.svg}{source}).}
    \label{CRwithHelper}
\end{figure}

Another extension of the two-source model is the helper-assisted setting~\cite{CRhelper}, illustrated in Fig.~\ref{CRwithHelper}. In this model, Alice and Bob generate CR with the assistance of a helper that communicates over a rate-limited noiseless public channel. The underlying discrete memoryless source produces i.i.d. samples of $(X,Y,Z)$, where the helper, Alice, and Bob observe $Z^n$, $X^n$, and $Y^n$, respectively. Communication takes place through two independent noiseless links: a helper-to-all link of rate $R_1$ and an Alice-to-Bob link of rate $R_2$. 

Compared with the one-way model, the helper-assisted setting introduces a third party with correlated side information. The helper communicates information about its observation, which Alice combines with her own observation to generate additional CR shared with Bob. This assistance can improve the achievable CR rate under the same communication constraints. The resulting achievable CR rate characterization is obtained by optimizing auxiliary RVs subject to the corresponding rate and Markov constraints~\cite{CRhelper}. \color{black}A matching converse is known only for certain special cases.\color{black}

\noindent\textbf{\color{black} Systems Perspective:\color{black}}
This model represents 6G scenarios in which relay nodes or edge servers assist secure coordination under communication constraints. 
\subsection{Two-Source Model for Interactive CR Generation over Independent DMCs}

\begin{figure}[H]
    \centering
    \input{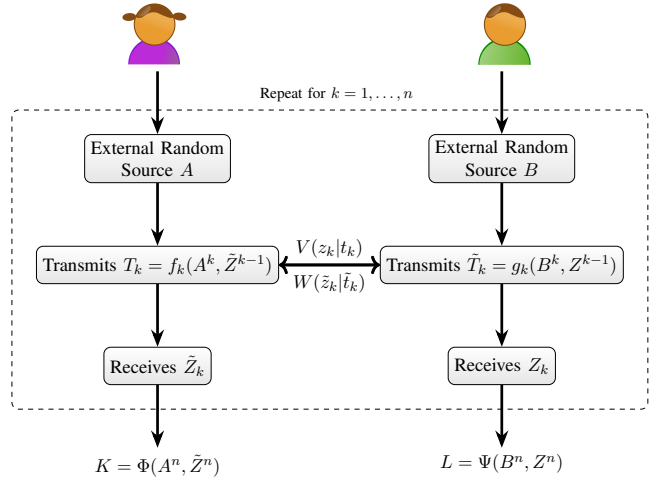}
    \caption{Protocol for interactive CR generation with external i.i.d.\ sources (\small Alice and Bob icons: see attribution in Fig. \ref{jammer}).}
    \label{CRpairfig}
\end{figure}

The problem of interactive CR generation, shown in Fig.~\ref{CRpairfig}, was introduced in \cite{CRpair}. 
Alice and Bob observe i.i.d.\ sources $\{A_k\}_{k=1}^\infty$ and $\{B_k\}_{k=1}^\infty$ with entropies $H_A$ and $H_B$ bits per symbol, respectively. Over $n$ steps, they exchange
$T_k = f_k(A^k, \tilde{Z}^{k-1})$ and 
$\tilde{T}_k = g_k(B^k, Z^{k-1})$,
where $Z^{k-1}$ and $\tilde{Z}^{k-1}$ denote past received messages, and generate $K = \Phi(A^n, \tilde{Z}^n)$ and $L = \Psi(B^n, Z^n)$, respectively. 

Compared with source-based CR generation, this model uses the communication channels as sources of randomness. The CR capacity is obtained by optimizing channel input distributions and is limited by two bottlenecks: transmitter-side randomness and reliable information transmission through the DMC 
\cite{CRpair}.

\noindent\textbf{\color{black} Systems Perspective:\color{black}}
This model is relevant to 6G systems that exploit both distributed device randomness and the inherent randomness of wireless channels, such as collaborative edge computing, MIMO random-access networks, and RIS-assisted communications.

\section{Application of CR Generation: CR-Assisted Secure Identification}
\label{application}

\begin{figure*}
    \centering
    \input{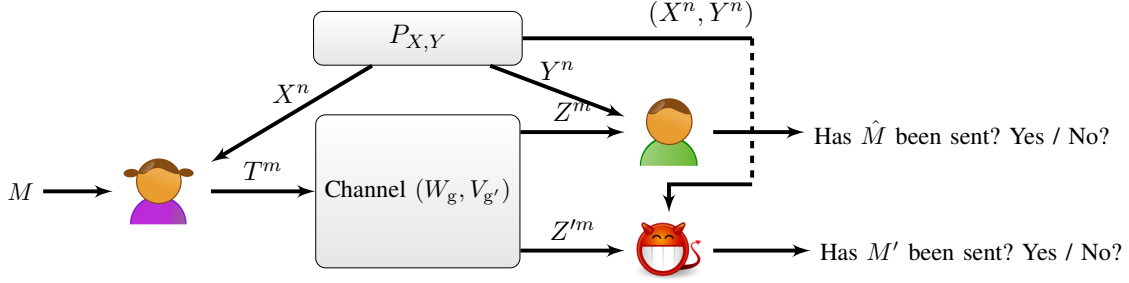}
    \caption{CR-assisted secure identification over the Gaussian wiretap channel (\small Alice and Bob icons: see attribution in Fig. \ref{jammer}; Eve icon: see attribution in Fig. \ref{seededmodularcoding}).}
    \label{fig:secureId}
\end{figure*}

\color{black}We consider
CR-assisted secure identification over the Gaussian wiretap channel (GWC), denoted by $(W_{\mathrm{g}}, V_{\mathrm{g'}})$ in Fig.~\ref{fig:secureId}, under an average power constraint $P$. The outputs at Bob and Eve are corrupted by i.i.d.\ Gaussian noise sequences, where each noise sample has variance $\sigma^2$ and $\sigma'^2,$ respectively, \color{black}and Eve has access to the correlated source outputs. \color{black}

\color{black}Returning to the V2V example, one possible scenario is that Vehicle A (Alice) broadcasts control messages over a wireless channel. Vehicle B (Bob) performs identification with respect to its own command of interest, such as a braking command, while an unauthorized vehicle (Eve) attempts to identify a different command, such as a lane-change instruction, using its channel observation and correlated side information. \color{black}

It is shown in \cite{optimalsignalprocessing} that, when the GWC secrecy capacity is positive, the availability of CR yields a secure identification capacity lower bounded by \(L_f^{X,Y}\!\left(C(g,P)\right)\), where \(C(g,P)\) denotes the main channel capacity. In contrast,  the secure identification capacity with only local randomness equals \(C(g,P)\). Therefore, when $C(g,P) \leq H(X),$ the availability of CR may enable secure identification rates beyond those achievable with local randomness.

\color{black}Next, we describe a coding scheme for secure identification \cite{optimalsignalprocessing}. \color{black} Alice computes a CR-based tag of the message and transmits it over the GWC using a wiretap code. Bob computes the tag of the message of interest using the CR and compares it with the received tag. Due to GWC secrecy, Eve cannot recover the tag and thus cannot reliably perform identification.\color{black}\color{black}
\begin{tcolorbox}
[
    title=Key Takeaway,
    colback=blue!5,
    colframe=blue!40!black,
    coltitle=white,
    colbacktitle=blue!40!black,
    boxrule=0.5pt,
    arc=1mm
]
The availability of CR may enable secure identification rates beyond those achievable with local randomness, providing significant performance gains.
\end{tcolorbox}

\begin{example}
\label{binarysources}
Suppose \(P_X(0)=P_X(1)=\frac{1}{2}\) and, for \(0\leq\mu\leq\frac{1}{2}\), define
\(P_{X,Y}(0,0)=P_{X,Y}(1,1)=\frac{1-\mu}{2}\) and
\(P_{X,Y}(0,1)=P_{X,Y}(1,0)=\frac{\mu}{2}\).
For a Gaussian channel to Bob with noise variance \(\sigma^2\), if
\[
P\geq P_{\star}=\sigma^{2}(2^{2H_{2}(\mu)}-1),
\]
then the CR capacity reaches its maximum \(H(X)=1\).
\color{black} Above this threshold, Alice and Bob can \color{black} establish \color{black} one bit of shared randomness per channel use, which is the maximum achievable CR rate, thereby enhancing identification performance. For $\mu=0.2$, $H_2(0.2)\approx0.7219$, yielding $P_\star\approx1.72\sigma^2$. With $\sigma^2=1$ and $P_\star \approx1.72$, the lower bound on the CR-assisted secure identification capacity reaches $1$, while the secure identification capacity with randomized encoding is equal to
\begin{align}
C(g,P_\star)
&=\frac{1}{2}\log_2\left(1+\frac{P_\star}{\sigma^2}\right) \approx0.72. \nonumber
\end{align}
Thus, the gain is approximately $0.28$. This comparison is illustrated in Fig.~\ref{figplotcomparetolowerbound}. From a system design perspective, this gain allows higher identification rates under the same power and secrecy constraints, improving secure communication efficiency in dense 6G deployments.
\begin{figure*}[hbt!]
\centering
    \hspace*{-0.48cm}
    \definecolor{mycolor1}{rgb}{0.00000,0.44700,0.74100}%
\definecolor{mycolor2}{rgb}{1.00000,0.00000,1.00000}%
\scalebox{.5792}{\begin{tikzpicture}

\begin{axis}[%
width=4.521in,
height=3.57in,
at={(0.758in,0.482in)},
scale only axis,
xmin=0,
xmax=3,
xlabel style={font=\color{white!15!black}},
xlabel={\textbf{Power}},
ymin=0,
ymax=1.01,
ylabel style={font=\color{white!15!black}},
ylabel={\textbf{Secure Identification Capacity}},
axis background/.style={fill=white},
xmajorgrids,
ymajorgrids,
legend style={at={(0.362,0.181)}, anchor=south west, legend cell align=left, align=left, draw=white!15!black}
]
\addplot [color=mycolor2, mark=square, mark options={solid, mycolor2}, line width=1.3pt]
  table[row sep=crcr]{%
0	0\\
0.4	0.3567159668\\
0.443700637652136	0.38698371754\\
0.487401275304271	0.42842710197\\
0.815858733228839   0.63181695969\\
0.865858733228839	0.65854911927\\
0.915858733228839	0.68388345766\\
1.25	0.83559452771\\
1.28947576084681	0.84376355186\\
1.32895152169362	0.85747865551\\
1.62047051030039	0.96316518122\\
1.67047051030039	0.97963138792\\
1.72047051030039	1\\
2	1\\
2.039600717839	1\\
2.079201435678	1\\
2.29304790375616	1\\
2.34304790375616	1\\
2.39304790375616	1\\
2.6	1\\
2.62530317679635	1\\
2.6506063535927	1\\
2.74211629784053	1\\
2.79211629784053	1\\
2.84211629784053	1\\
2.9	1\\
2.93006653339619	1\\
2.96013306679237	1\\
2.98006653339619	1\\
3	1\\
};
    \addlegendentry{\textbf{Lower bound on CR-assisted}\\\textbf{secure identification capacity}}

\addplot [color=mycolor1, line width=1.3pt]
  table[row sep=crcr]{%
0	0\\
0.01	0.00717764648853503\\
0.02	0.0142845760983855\\
0.03	0.0213221687042469\\
0.04	0.0282917641831838\\
0.05	0.035194663945699\\
0.06	0.0420321323942373\\
0.07	0.0488053983132112\\
0.08	0.055515656194372\\
0.09	0.0621640675011009\\
0.1	0.0687517618749675\\
0.11	0.0752798382876907\\
0.12	0.0817493661414398\\
0.13	0.0881613863202314\\
0.14	0.0945169121950086\\
0.15	0.100816930584825\\
0.16	0.107062402676424\\
0.17	0.11325426490434\\
0.18	0.119393429793558\\
0.19	0.125480786766609\\
0.2	0.131517202916897\\
0.21	0.137503523749935\\
0.22	0.143440573894081\\
0.23	0.149329157782258\\
0.24	0.155170060306075\\
0.25	0.160964047443681\\
0.26	0.166711866862596\\
0.27	0.172414248498721\\
0.28	0.178071905112638\\
0.29	0.183685532824265\\
0.3	0.189255811626865\\
0.31	0.194783405881363\\
0.32	0.200268964791864\\
0.33	0.205713122863233\\
0.34	0.211116500341524\\
0.35	0.216479703638053\\
0.36	0.221803325737807\\
0.37	0.227087946592901\\
0.38	0.232334133501722\\
0.39	0.237542441474391\\
0.4	0.242713413585121\\
0.41	0.247847581312035\\
0.42	0.252945464864979\\
0.43	0.258007573501832\\
0.44	0.263034405833794\\
0.45	0.268026450120105\\
0.46	0.272984184552646\\
0.47	0.27790807753082\\
0.48	0.282798587927113\\
0.49	0.287656165343718\\
0.5	0.292481250360578\\
0.51	0.297274274775177\\
0.52	0.30203566183443\\
0.53	0.306765826458964\\
0.54	0.311465175460088\\
0.55	0.316134107749756\\
0.56	0.320773014543762\\
0.57	0.325382279558451\\
0.58	0.329962279201189\\
0.59	0.334513382754815\\
0.6	0.339035952556319\\
0.61	0.343530344169946\\
0.62	0.34799690655495\\
0.63	0.352435982228176\\
0.64	0.35684790742168\\
0.65	0.361233012235545\\
0.66	0.3655916207861\\
0.67	0.369924051349664\\
0.68	0.374230616502018\\
0.69	0.37851162325373\\
0.7	0.382767373181489\\
0.71	0.386998162555587\\
0.72	0.391204282463687\\
0.73	0.395386018931\\
0.74	0.399543653037002\\
0.75	0.403677461028802\\
0.76	0.407787714431286\\
0.77	0.411874680154136\\
0.78	0.415938620595837\\
0.79	0.419979793744766\\
0.8	0.423998453277475\\
0.81	0.42799484865424\\
0.82	0.431969225211986\\
0.83	0.435921824254659\\
0.84	0.439852883141144\\
0.85	0.443762635370794\\
0.86	0.447651310666653\\
0.87	0.451519135056456\\
0.88	0.455366330951456\\
0.89	0.459193117223174\\
0.9	0.462999709278112\\
0.91	0.466786319130512\\
0.92	0.470553155473216\\
0.93	0.474300423746678\\
0.94	0.478028326206201\\
0.95	0.481737061987443\\
0.96	0.485426827170242\\
0.97	0.489097814840826\\
0.98	0.492750215152442\\
0.99	0.496384215384462\\
1	0.5\\
1.01	0.503597750702102\\
1.02	0.507177646488535\\
1.03	0.510739863705226\\
1.04	0.514284576098385\\
1.05	0.517811954865361\\
1.06	0.521322168704247\\
1.07	0.5248153838623\\
1.08	0.528291764183184\\
1.09	0.531751471153079\\
1.1	0.535194663945699\\
1.11	0.53862149946623\\
1.12	0.542032132394237\\
1.13	0.545426715225557\\
1.14	0.548805398313211\\
1.15	0.552168329907368\\
1.16	0.555515656194372\\
1.17	0.558847521334877\\
1.18	0.562164067501101\\
1.19	0.565465434913224\\
1.2	0.568751761874967\\
1.21	0.572023184808353\\
1.22	0.575279838287691\\
1.23	0.57852185507279\\
1.24	0.58174936614144\\
1.25	0.584962500721156\\
1.26	0.588161386320232\\
1.27	0.591346148758095\\
1.28	0.594516912195009\\
1.29	0.59767379916111\\
1.3	0.600816930584825\\
1.31	0.603946425820666\\
1.32	0.607062402676424\\
1.33	0.610164977439778\\
1.34	0.61325426490434\\
1.35	0.616330378395137\\
1.36	0.619393429793558\\
1.37	0.622443529561767\\
1.38	0.625480786766609\\
1.39	0.628505309103012\\
1.4	0.631517202916897\\
1.41	0.634516573227619\\
1.42	0.637503523749935\\
1.43	0.640478156915528\\
1.44	0.643440573894081\\
1.45	0.646390874613923\\
1.46	0.649329157782258\\
1.47	0.652255520904976\\
1.48	0.655170060306075\\
1.49	0.658072871146678\\
1.5	0.660964047443681\\
1.51	0.663843682088024\\
1.52	0.666711866862596\\
1.53	0.669568692459793\\
1.54	0.672414248498721\\
1.55	0.675248623542066\\
1.56	0.678071905112638\\
1.57	0.680884179709577\\
1.58	0.683685532824265\\
1.59	0.686476048955915\\
1.6	0.689255811626865\\
1.61	0.69202490339758\\
1.62	0.694783405881363\\
1.63	0.697531399758789\\
1.64	0.700268964791864\\
1.65	0.702996179837918\\
1.66	0.705713122863233\\
1.67	0.708419870956415\\
1.68	0.711116500341524\\
1.69	0.71380308639095\\
1.7	0.716479703638053\\
1.71	0.719146425789573\\
1.72	0.721803325737807\\
1.73	0.724450475572564\\
1.74	0.727087946592901\\
1.75	0.729715809318649\\
1.76	0.732334133501722\\
1.77	0.734942988137232\\
1.78	0.737542441474391\\
1.79	0.740132561027231\\
1.8	0.742713413585121\\
1.81	0.745285065223101\\
1.82	0.747847581312034\\
1.83	0.750401026528579\\
1.84	0.752945464864979\\
1.85	0.75548095963869\\
1.86	0.758007573501832\\
1.87	0.760525368450482\\
1.88	0.763034405833794\\
1.89	0.765534746362977\\
1.9	0.768026450120105\\
1.91	0.77050957656678\\
1.92	0.772984184552646\\
1.93	0.775450332323761\\
1.94	0.77790807753082\\
1.95	0.780357477237239\\
1.96	0.782798587927113\\
1.97	0.78523146551302\\
1.98	0.787656165343718\\
1.99	0.79007274221169\\
2	0.792481250360578\\
2.01	0.794881743492489\\
2.02	0.797274274775177\\
2.03	0.799658896849113\\
2.04	0.80203566183443\\
2.05	0.804404621337762\\
2.06	0.806765826458964\\
2.07	0.809119327797727\\
2.08	0.811465175460088\\
2.09	0.813803419064825\\
2.1	0.816134107749756\\
2.11	0.818457290177939\\
2.12	0.820773014543762\\
2.13	0.823081328578947\\
2.14	0.825382279558451\\
2.15	0.827675914306277\\
2.16	0.829962279201189\\
2.17	0.832241420182341\\
2.18	0.834513382754815\\
2.19	0.836778211995072\\
2.2	0.839035952556319\\
2.21	0.841286648673789\\
2.22	0.843530344169946\\
2.23	0.8457670824596\\
2.24	0.84799690655495\\
2.25	0.850219859070546\\
2.26	0.852435982228176\\
2.27	0.854645317861679\\
2.28	0.856847907421679\\
2.29	0.859043791980258\\
2.3	0.861233012235545\\
2.31	0.863415608516247\\
2.32	0.8655916207861\\
2.33	0.867761088648269\\
2.34	0.869924051349664\\
2.35	0.872080547785205\\
2.36	0.874230616502018\\
2.37	0.876374295703567\\
2.38	0.87851162325373\\
2.39	0.88064263668081\\
2.4	0.882767373181488\\
2.41	0.884885869624724\\
2.42	0.886998162555587\\
2.43	0.889104288199044\\
2.44	0.891204282463687\\
2.45	0.893298180945403\\
2.46	0.895386018931\\
2.47	0.897467831401768\\
2.48	0.899543653037002\\
2.49	0.901613518217464\\
2.5	0.903677461028802\\
2.51	0.905735515264918\\
2.52	0.907787714431286\\
2.53	0.909834091748228\\
2.54	0.911874680154136\\
2.55	0.91390951230866\\
2.56	0.915938620595837\\
2.57	0.917962037127187\\
2.58	0.919979793744766\\
2.59	0.921991922024163\\
2.6	0.923998453277475\\
2.61	0.925999418556223\\
2.62	0.92799484865424\\
2.63	0.929984774110513\\
2.64	0.931969225211986\\
2.65	0.933948231996327\\
2.66	0.935921824254659\\
2.67	0.937890031534244\\
2.68	0.939852883141144\\
2.69	0.941810408142836\\
2.7	0.943762635370794\\
2.71	0.945709593423039\\
2.72	0.947651310666653\\
2.73	0.949587815240257\\
2.74	0.951519135056456\\
2.75	0.953445297804259\\
2.76	0.955366330951456\\
2.77	0.95728226174697\\
2.78	0.959193117223174\\
2.79	0.961098924198184\\
2.8	0.962999709278112\\
2.81	0.964895498859299\\
2.82	0.966786319130512\\
2.83	0.968672196075116\\
2.84	0.970553155473216\\
2.85	0.97242922290377\\
2.86	0.974300423746678\\
2.87	0.976166783184843\\
2.88	0.978028326206202\\
2.89	0.979885077605734\\
2.9	0.981737061987443\\
2.91	0.983584303766314\\
2.92	0.985426827170242\\
2.93	0.987264656241941\\
2.94	0.989097814840826\\
2.95	0.99092632664487\\
2.96	0.992750215152442\\
2.97	0.994569503684117\\
2.98	0.996384215384462\\
2.99	0.998194373223811\\
3	1\\
};
\addlegendentry{\textbf{Capacity of secure identification}\\  \textbf{with randomized encoding}}

\end{axis}
\begin{axis}[%
width=5.833in,
height=4.375in,
at={(0in,0in)},
scale only axis,
xmin=0,
xmax=1,
ymin=0,
ymax=1,
axis line style={draw=none},
ticks=none,
axis x line*=bottom,
axis y line*=left,
legend style={legend cell align=left, align=left, draw=white!15!black}
]
\end{axis}
\end{tikzpicture}}%
    \caption{Comparison of the lower bound on CR-assisted secure identification capacity and secure identification capacity with randomized encoding for \(\sigma^2=1\) and \(\mu=0.2\).}
    \label{figplotcomparetolowerbound}
\end{figure*}
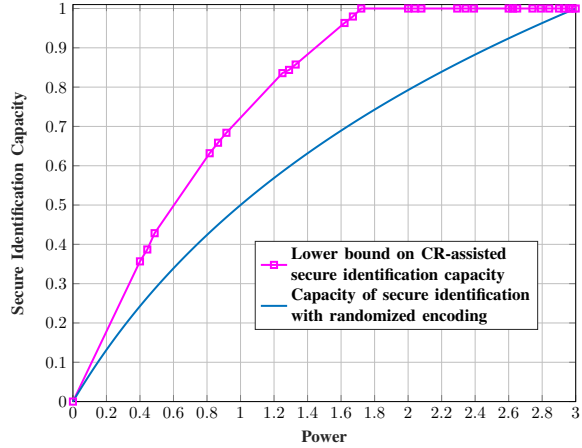

\end{example}

\section{Open Information-Theoretic Problems}
\color{black}Despite significant progress in CR generation theory, several information-theoretic challenges remain for future 6G systems. We highlight three key directions below: \color{black}
\setlength{\emergencystretch}{2em}
\begin{itemize}
\item \textbf{Beyond i.i.d. Setting:} \color{black}A common assumption in CR generation is that the parties observe i.i.d.\ samples of a correlated source. To the best of our knowledge, a general characterization for non-stationary and time-varying sources is still lacking. This is relevant to 6G systems, where mobility and environmental dynamics can lead to time-varying source statistics. \color{black}
\item \textbf{Finite-Length Analysis:} \color{black}Most CR generation results are asymptotic. Although some works have studied finite-blocklength CR generation\cite{nonasymptotic,secondorder}, a comprehensive characterization at practical blocklengths remains challenging. This is important for 6G services such as URLLC, where communication occurs over short packets. \color{black} 
\item\textbf{Multi-Party Setting:} Although CR generation has been explored in multi-party settings in some prior work~\cite{CRnetwork}, its extension to more general multi-party network models remains challenging.
This is relevant to
6G architectures such as cooperative V2X,
edge intelligence, and cell-free systems.\color{black}%
\end{itemize}
\color{black}
\section{Acknowledgments} \color{black}The authors acknowledge support from the BMFTR under the “Souverän. Digital. Vernetzt.” programme for project 6G-life (16KIS2414, 16KIS2415). H. Boche and C. Deppe acknowledge support from the BMFTR Quantum Programme Q-STARS (16KIS2601, 16KIS2602) and DFG support under the project "Post Shannon Theorie und Implementierung" (BO 1734/38-1, DE 1915/2-1). H. Boche and W. Labidi received BMFTR support through the Internet of Bio-Nano-Things (IoBNT) via Grant 5310223.
\color{black}H. Boche acknowledges funding by the DFG as part of Germany’s Excellence Strategy-EXC 2050/2-Project ID 390696704-Cluster of Excellence ”Centre for
Tactile Internet with Human-in-the-Loop” (CeTI) of Technische Universit\"at
Dresden. \color{black}
M. Wiese received support from the Bavarian Ministry of Economic Affairs, Regional Development and Energy for the 6G Future Lab Bavaria project. \color{black}
\section{Author Bios}
 \textbf{Rami Ezzine} (Graduate Student Member, IEEE) 
received the B.Sc. and M.Sc. degrees in electrical engineering from TU Munich (TUM), Germany, in 2016 and 2019, respectively. He is currently pursuing the Ph.D. degree at the Chair of Theoretical Information Technology, TUM, under the supervision of Prof. Boche. His research interests include information theory and Post-Shannon theory, with a particular focus on CR  generation.

\textbf{Moritz Wiese} (Member, IEEE) received the Dipl.-Math. degree in mathematics from
the University of Bonn, Germany, in 2007. He obtained the PhD
degree from TUM in 
2013. From 2007 to 2010, he was a research assistant at TU Berlin, Germany, and from 2010 until 2014 at TUM. From 2014 to 2016, he was with the ACCESS 
Linnaeus Center at KTH Royal Institute of Technology, Stockholm,
Sweden. Currently, he is with the Chair of Theoretical Information
Technology at TUM.

\textbf{Wafa Labidi} (Graduate Student Member, IEEE) 
received the B.Sc. and M.Sc. degrees in electrical engineering from TUM, Germany, in 
2016 and 2019, respectively. She is currently working toward the
Ph.D. degree at the Chair of Theoretical Information Technology,
TUM. She is supervised by Prof. Boche. \color{black}Her research mainly focuses on message identification, CR generation, and molecular
communication. \color{black}

\textbf{Christian Deppe} (Senior Member, IEEE) received the Dipl.-Math. degree
in mathematics from the Universit\"at Bielefeld, Germany, 
in 1996, and the Dr.-Math. degree in mathematics from the Universit\"at Bielefeld, Germany, in 1998. He was a Research and
Teaching Assistant with the Fakult\"at f\"ur Mathematik, Universit\"at
Bielefeld, from 1998 to 2017. \color{black}He spent six years at the TUM Chair of Communications Engineering before joining TU Braunschweig in January 2024. \color{black} His research interests include Post-Shannon Theory, Quantum Communication Networks, and Molecular Communication.

\textbf{Holger Boche} (Fellow, IEEE) received the Dipl.-Ing. degree in
electrical engineering, the Graduate degree in mathematics, and
the Dr.-Ing. degree in electrical engineering from TU Dresden, Germany, in 1990, 1992, and 1994, and the Dr. rer. nat. degree in 
pure mathematics from the TU Berlin, Germany, in 1998. He is currently a
Full Professor at the Chair of Theoretical Information Technology,
TUM, which he 
joined in October 2010. He is head of BMFTR Research Hub 6G-life. Its central goal is to lay the theoretical foundations for 6G communication systems.

\end{document}